\documentstyle{article}
\def\Wt{\tilde{W}}
\def\wt{\tilde{w}}
\def\R{{\cal R}}
\def\sg{\sigma}
\def\beq{\begin{equation}}
\def\eeq{\end{equation}}
\def\bey{\begin{eqnarray}}
\def\eey{\end{eqnarray}}
\begin{document}
\baselineskip .2in
\begin{flushright}
PUPT-1290  \\
November 1991 \\
\end{flushright}
\vspace{20mm}
\begin{center}
{\large ON LOOP EQUATIONS IN KdV \\
EXACTLY SOLVABLE STRING THEORY\\}
\vspace{30mm}
{\bf Simon Dalley} \\
{\em Joseph Henry Laboratories, Princeton University, \\
Princeton, New Jersey 08544, U.S.A.\\}
\vspace{10mm}
{\bf Abstract}\\
\end{center}
\noindent The non-perturbative
behaviour of macroscopic loop amplitudes in the exactly solvable string
theories based on the KdV hierarchies is considered.
Loop equations are presented for the real non-perturbative
solutions living on the spectral half-line, allowed by the most general
string equation $[\tilde{P},Q]=Q$, where $\tilde{P}$ generates scale
transformations. In general the end of the half-line (the `wall') is a
non-perturbative parameter whose r\^ole is that of boundary cosmological
constant. The properties are compared with the perturbative behaviour and
solutions of $[P,Q]=1$. Detailed arguments are given for
the $(2,2m-1)$  models while generalisation to the other $(p,q)$ minimal
models and $c=1$ is briefly addressed.
\vfill
\newpage
\section{Introduction}
Soon after the seminal works on non-perturbative 2D gravity appeared
\cite{new1}, a number of authors discussed the behaviour of macroscopic
string amplitudes in those theories \cite{bdss,bmp,dav}. In genus
perturbation theory these are the surfaces with boundaries whose lengths
scale in the continuum limit. It is possible however to discuss such
correlators independently of this weak coupling expansion. Since the latter
is only asymptotic and typically not Borel resummable one must input some
non-perturbative information in order to even begin discussing such
questions. The mathematical possibilities are of course in principle endless.
It is not so easy however to come up with a systematic formalism. In what
follows the extra information that is required will be taken to be the
principle that the KdV structure organising genus perturbation theory is
present non-perturbatively\footnote{This choice clearly has aesthetic and
computational qualities but is nevertheless not, at present, a scientific
argument.}. This point of view was taken to its logical conclusion in
refs.\cite{djm1,djm2} where it was found that new non-perturbative solutions
were allowed that had not been previously considered. It is expedient to
briefly review the main points of that work. Consider the $(2,2m-1)$
minimal models, for which one has the KdV flows
$\partial_{t_{k}}u=-\partial_{t_{0}}R_{k+1}[u]$
(notation is hopefully standard), which define the
isospectral symmetries $u \rightarrow u+\epsilon\partial_{t_{k}}u$
leaving invariant the spectrum of the hamiltonian operator
$-\partial^{2}_{t_{0}} + u$. Once the string
susceptibility $u$ has been determined these flows yield
correlators of the local operators with couplings $t_{k}$. The minimal physical
requirement that one can impose on $u$
is its scaling equation under a change of
the length scale in the theory. For the set of parameters
$\{u,\{{t}_{k}\}\}$ the
KdV flows determine this equation to be \cite{djm1};
\beq
L\R = 0 \label{smiley}
\eeq
where
\bey
\R & = & \sum_{k\geq 0} (k+1/2){t}_{k}R_{k} \nonumber \\
& = & {t}_{0} - \sum_{k\geq 1}(k+1/2){t}_{k}(D^{-1}L)^{k-1}u \\
D=\nu\frac{\partial}{\partial {t}_{0}} &,&
L=-\frac{1}{4}D^{3} +\frac{1}{2} uD +
\frac{1}{2} Du
\eey
and $\nu$ is the renormalised value of $1/N$.
The string equation (\ref{smiley}) may be
written as the canonical commutation relation on the half-line, $[\tilde{P},Q]
=Q$, for co-ordinate (Lax) operator $Q=D^{2}-u$ and $\tilde{P}=\sum_{k\geq 0}
t_{k} Q_{+}^{k+1/2}$ the generator of
scale transformations of these co-ordinates, in a double-scaled Dyson gas
with polynomial potential.

To reproduce the $\nu$-perturbative results as specified by
the matrix models one must choose a solution which is asymptotically a solution
of $\R =0$ at $t_{0} \rightarrow -\infty$. One possibility is to choose an
exact solution of $\R =0$, for which there are two types of real solution
$u$. Those accumulating poles as $t_{0} \rightarrow \infty$ \cite{new1} (type
1), and solutions of $k$ odd critical points and non-singular flows around
those points \cite{bmp} (type 2), for which $u$ has a real asymptotic
expansion at $t_{0} \rightarrow \infty$.
There is one other type of real solution to (\ref{smiley}) with correct
asymptotic behaviour as $t_{0} \rightarrow -\infty$ (type 3). This has no
poles\footnote{The pole-free nature of
the solutions was verified for the $k=1,2$ critical points in ref.\cite{djm1}
and has also been checked for $k=3$ and the Ising model
\cite{bill}} and has $u \rightarrow 0$ as $t_{0} \rightarrow \infty$
($\R \rightarrow {t}_{0}$).
It is realised by a Dyson gas restricted to lie on the spectral half-line
i.e. there is an infinite potential `wall' at the critical
edge of the charge distribution. In the following section this representation
will be used to derive Dyson-Schwinger loop equations.
 Figure 1 shows the numerical result for
the type 3 susceptibility for pure gravity found in ref.\cite{djm1}.
\section{Type 3 Loop equations}
The scaling symmetry that led to (\ref{smiley}) is one of the non-isospectral
symmetries of the KdV hierarchy (see e.g.\cite{la}).
Together with the isospectral
symmetries (KdV flows) it implies a family of constraint equations by applying
the recursion operator $LD^{-1}$;
\beq
(LD^{-1})^{n} L\R = 0 \;\;\; , \;\;\; n\geq 0
\eeq
These take the form of Virasoro constraints $L_{n} \tau = 0$,
$u=-2D^{2}\log{\tau}$ \cite{vir}. There is an additional non-isospectral
symmetry which, if applied, constrains the solution $u$ further. Invariance
under this transformation is the $L_{-1}$ constraint $D{\R}=0$,
which is the hermitian matrix model string equation.
 This is the Galilean transformation: A
translation of the Dyson gas potential, given in infinitesimal form by;
\bey
u & \longrightarrow & \tilde{u} = u - \epsilon \nonumber \\
t_{k} & \longrightarrow & \tilde{t}_{k} = t_{k} +
\epsilon (k+3/2) t_{k+1} \label{gal}
\eey
is an invariance of $\R$ but is not respected by (\ref{smiley}).

In the case of type 3 there is a further parameter in the theory
which has so far
been neglected \cite{djm2}. For the solution on the half-line one should allow
the infinite potential wall to be at some arbitrary scaled position, $\sg$ say.
This modifies the canonical position and momentum to $Q+\sg$ and $\tilde{P}
+\sg P$, where $P= \sum_{k\geq 1}t_{k}Q_{+}^{k-1/2}$
is the translation operator.
In this way the string equation
is modified to \cite{djm2};
\beq
[\tilde{P} + \sg P, Q + \sg ] = Q + \sg
\eeq
equivalently
\beq
L\R - \sg D\R = 0 \label{newsm}
\eeq
Since $\sg$ has the same dimension as $u$ and (\ref{newsm}) is the scaling
equation one identifies the evolution with respect to $\sg$:
$\nu\partial_{\sg} u = -D\R$.
This is the differentiated form of the $L_{-1}$ condition in the present case:
$L_{-1} \tau = \partial_{\sg} \tau$.
The Virasoro constraints are the expression of diffeomorphism invariance of the
spectral line and the presence of a `wall' has induced a boundary term on the
right hand side. Similar boundary terms appear in the other
constraints when $\sg \neq 0$, since varying the boundary as $\sg \rightarrow
\sg + \epsilon {\sg}^{n+1}$ implies that
\beq
L_{n} \tau = {\sg}^{n+1} \frac{\partial \tau}{\partial \sg} \label{len}
\eeq
The Galilean and higher KdV symmetries are now respected when the correct
transformation properties of $\sg$ are taken into account. In particular to
(\ref{gal}) one must add $\sg \longrightarrow \sg - \epsilon$.
(\ref{newsm}) becomes an equation for $\tilde{u}$ in terms of $\tilde{t}$,
independent of $\sg$.
The constraints (\ref{len}) may be rewritten in the old form
for $n \geq 0$ by using the
invariance of the Virasoro algebra under
\beq
L_{n} \longrightarrow \tilde{L}_{n} = {\rm e}^{-\sg L_{-1}} L_{n}
{\rm e}^{\sg L_{-1}} \label{fgal}
\eeq
Given $L_{-1}\tau = \partial_{\sg} \tau$ the equations (\ref{len}) are easily
seen to be equivalent to $\tilde{L}_{n} \tau =0$.
The transformation (\ref{fgal})
describes a {\em finite} Galilean transformation by $\sg$, and the
possibility of making redefinitions of this sort can be
 viewed as another reason for
introducing $\sg$ in the most general framework.

The parameter $\sg$ transforms in the same way as $t_{m-1}/t_{m}$ under
(\ref{gal}), where $t_{k}=0$, $k>m$ and $t_{m}$ is invariant in this case.
Following the reasoning of ref.\cite{mms} it suggests that  its r\^ole is
 that of boundary
cosmological constant. Indeed if one takes the usual expression
for the renormalised
macroscopic loop wavefunction \cite{bdss};
\beq
<\!w(l)\!>  =   \int_{-t_{0}}^{\infty} <\! z | {\rm exp}l(\nu^{2}
\partial_{z}^{2} -u) |z \!> dz \label{loo}
\eeq
a finite Galilean transformation by $\sg$ exhibits the ${\rm e}^{-\sg l}$
dependence of the loop. The net effect of the transformation is to define
away the non-exponential dependence of $<\!w(l)\!>$ upon $\sg$.
In this case one can now say that $\sg$ couples to
the boundary operator \cite{mms} in the sense that
\beq
\left(\frac{\partial}{\partial \sg}\right)_{\tilde{t}}
 <\!\prod_{i} w(l_{i})\!> = \sum_{i}l_{i}
<\! \prod_{i} w(l_{i}) \!>
\eeq
where
\bey
\left(\frac{\partial}{\partial \sg}\right)_{\tilde{t}} & = &
\left(\frac{\partial}{\partial \sg}\right)_{t} - \sum_{k\geq 1} (k+1/2)t_{k}
\frac{\partial}{\partial t_{k-1}}  \\
\left(\frac{\partial^{n} u}{\partial \sg^{n}} \right)_{\tilde{t}} & = &
\delta_{n1}
\eey

It is worth noting that there is a certain amount of ambiguity in the
definition of the loop wavefunction, the choice (\ref{loo}) being
perhaps the simplest. Two reasonable requirements might be
that it agree with the results of Liouville theory and have a local (KdV)
operator expansion at small $l$. The former is meaningful perturbatively
in $\nu$ and the latter perturbatively in $l$. This does not fix
contributions non-perturbative in both of these parameters however. In
this letter the definition (\ref{loo}) will always be assumed.

The arguments up to now have been somewhat heuristic so it is instructive
to give a more careful treatment of some of the points. First a proof of
the fact that (\ref{newsm}) is the string equation of a Dyson gas on
$[\sg,\infty)$ will be given. To agree with the
conventions for non-universal constants adopted implicitly earlier,
the following calculation is performed with an even polynomial
action $NV/\Lambda$ on the interval
$[-2,2]$, the ends of the charge distribution coinciding with infinite walls
at $\pm 2$ (in un-scaled variables). The scaling regions around $\pm 2$ thus
furnish identical copies of the system on the half-line.
Introducing an infinitesimal cutoff $\delta$, let the scaled positions of the
walls be $\pm 2 \mp\sg\delta^{2}$. Further renormalised parameters are defined
by;
\bey
\Lambda  = & 1+ t_{0}\delta^{2m} \;\;\;\; & \frac{1}{N} = \nu
\delta^{2m+1} \nonumber \\
\frac{\Lambda n}{N} = &  1-z\delta^{2m} \;\;\;\; &
 R_{n}=1-u(z)\delta^{2}, \; n \sim N
\eey
for the neighbourhood of the $m$th critical point. The equations of motion in
the orthogonal polynomial formalism \cite{biz} are;
\bey
\frac{(2n+1)\Lambda}{N} - <\!n|\lambda V^{'} |n \!> & = & \frac{2\Lambda}{N}
(2-\sg\delta^{2})P_{n}^{2} \label{emot1} \\
\frac{n\Lambda}{N} - \sqrt{R_{n}}\! <\!n-1|V^{'}|n\!> & = &
\frac{2\Lambda}{N} \sqrt{R_{n}} P_{n} P_{n-1} \label{emot2}
\eey
The notation $P_{n}$ is shorthand for $P_{n}(2){\rm exp}(-NV(2)/2\Lambda)$
where $P_{n}(\lambda)$ are polynomials orthogonal on $[-2,2]$. Terms
involving $P_{n}$ in (\ref{emot1})(\ref{emot2}) are boundary terms picked up
when one integrates by parts standard identities to derive the equations of
motion. The left hand side of (\ref{emot1}) is;
\bey
&& \frac{(2n+1)\Lambda}{N} - \sqrt{R_{n+1}}\!<\! n | V^{'}|n+1\!> -
\sqrt{R_{n}} \! <\! n-1|V^{'}|n \!> \label{c} \\
& = & \frac{2\Lambda}{N} (\sqrt{R_{n+1}} P_{n} P_{n+1} + \sqrt{R_{n}} P_{n}
P_{n-1} ) \label{d}
\eey
using (\ref{emot2}). From the work on polynomial potentials \cite{new1};
\beq
\frac{n\Lambda}{N} - <\! n-1 |V^{'} |n \!>\!\sqrt{R_{n}} =
\delta^{2m} (\R (z) + O(\delta^{2}) ) \label{gm}
\eeq
{}From (\ref{emot1})(\ref{c}) and (\ref{gm}) one has;
\bey
\delta^{2m} Y^{2} & = & \frac{2\Lambda}{N}(2-\sg\delta^{2})P_{n}^{2} \\
Y & = & (\R (z) + \R (z+\nu\delta) + O(\delta^{2}))^{1/2}
\eey
and eliminating $P_{n},P_{n-1}$ from (\ref{d}) gives the string equation
at the first non-trivial order in $\delta$;
\beq
\frac{\nu^{2}}{2}\R\R^{''} - \frac{\nu^{2}}{4}(\R^{'})^{2} -
(u-\sg)\R^{2} = 0 \label{box}
\eeq
Differentiating once yields (\ref{newsm}). The Dyson gas also supplies
the appropriate boundary conditions $u \rightarrow \sg$ as $t_{0}
\rightarrow \infty$ since $u$ marks the edge of the charge distribution
in the leading WKB approximation. In fact a BIPZ analysis \cite{bipz,djm1}
shows that in this approximation the charge density aquires a square-root
divergence at the wall ($t_{0} >0$). More generally at $t_{0}<0$ this
divergence has exponentially small residue, as can be most easily seen from
the conventional form of the Dyson-Schwinger equations. These can be derived
in the usual way \cite{wad} since the type 3 solutions have a path integral
representation, and should correspond to the Virasoro constraints
described earlier when Taylor expanded along the lines of refs.\cite{vir}.
Starting from the partition function on $[-\Sigma,\Sigma]$ say,
in unscaled variables;
\beq
Z = \int_{-\Sigma}^{\Sigma} \prod_{i=1}^{N} d\lambda_{i} \;
\Delta^{2}(\lambda){\rm e}^{-N
\sum_{i} V(\lambda_{i})} \label{par}
\eeq
and defining the loop generating function and its
Laplace transform (marked loops are
used for convenience);
\bey
W(L) & = & \frac{1}{N}\sum_{i} {\rm e}^{L\lambda_{i}} \\
\Wt (T) & = & \frac{1}{N} \sum_{i} \frac{1}{T-\lambda_{i}}
\eey
one may perform an infinitesimal change of variables $\lambda_{i} \rightarrow
\lambda_{i} + \epsilon/(T-\lambda_{i})$ in (\ref{par}). The only new
contribution to the standard analysis \cite{wad} is from the variation of the
boundaries, giving
\beq
\epsilon \frac{\partial Z}{\partial \Sigma} \left( \frac{1}{T-\Sigma} -
\frac{1}{T+\Sigma} \right)
\eeq
The first loop equation is then
\beq
V^{'}(T)<\!\Wt (T) \!> + \Pi (T) = <\!\Wt (T) \Wt (T) \!> +
\frac{2\Sigma}{N^{2}(T^{2}-\Sigma^{2})} \frac{\partial Z}{\partial \Sigma}
\label{dag}
\eeq
where $\Pi (T)$ is a linear combination of $<\! W(0) \!>$,$<\! W^{'}(0) \!>$,
$\ldots$. In terms of $W(L)$ one has the equation of motion (introducing
connected correlators);
\bey
V^{'}\left(\frac{\partial}{\partial L}\right) <\! W(L) \!>_{c} & = &
\int_{0}^{L}dL^{'} \; (\frac{1}{N^{2}}<\! W(L^{'}) W(L-L^{'}) \!>_{c}
\nonumber \\
&+ & <\! W(L^{'})\!>_{c}<\! W(L-L^{'})\!>_{c})  -
\frac{2{\rm sinh}L\Sigma}{N^{2}}
\frac{\partial \log{Z}}{\partial \Sigma}
\eey
showing a new source-like term for the loop generating function.

Following ref.\cite{dav}
it is a simple matter to take the continuum limit by introducing, in the
neighbourhood of the $m$th critical point, renormalised parameters;
\bey
\Sigma = \Sigma_{c} - \sg \delta^{2} & ,& T= -\Sigma_{c} - \tau \delta^{2}
\nonumber \\
<\!\Wt (T) \!>_{c} & = & \frac{1}{2}V^{'}(T) + \delta^{2m-1}<\!\wt
(\tau)\!>_{c} \\
<\!\prod_{i=1}^{M} \Wt (T_{i})  \!>_{c} & = & \delta^{4m+2-M(2m+3)}
<\!\prod_{i=1}^{M} \wt (\tau_{i}) >_{c} \nonumber
\eey
Equation (\ref{dag}) becomes
\beq
<\! \wt (\tau) \!>^{2}_{c} + \nu^{2}\!<\! \wt (\tau) \wt (\tau) \!>_{c} =
\pi (\tau) + \frac{\nu^{2}}{\sg + \tau} \frac{\partial \log{Z}}{\partial
\sg} \label{star}
\eeq
For completeness the higher loop equations are
\bey
 && 2<\! \wt (\tau) \!>_{c}<\! \wt (\tau) \prod_{i=1}^{M} \wt (\tau_{i})
\!>_{c} + \nu^{2}\!<\! \wt (\tau) \wt (\tau) \prod_{i=1}^{M} \wt (\tau_{i})
\!>_{c} \nonumber \\
&& + \sum_{I,J} <\! \wt (\tau) \prod_{i\epsilon I} \wt (\tau_{i}) \!>_{c}
<\! \wt (\tau) \prod_{j\epsilon J} \wt (\tau_{j}) \!>_{c} \nonumber \\
&& + \sum_{i=1}^{M} <\!\wt (\tau_{1}) \ldots \frac{\partial}{\partial
\tau_{i}} \frac{\wt (\tau_{i}) - \wt (\tau)}{\tau_{i} - \tau} \ldots
\wt (\tau_{M}) \!>_{c} \nonumber \\
&& = <\! O_{0} \prod_{i=1}^{M} \wt (\tau_{i}) \!>_{c} + \frac{\nu^{2}}{
\sg + \tau} \frac{\partial}{\partial \sg} \!<\! \prod_{i=1}^{M} \wt
(\tau_{i}) \!>_{c}
\eey
where $O_{0}$ is the (bulk) puncture operator and the index sets $I,J$ are
such that $I \cup J = \{1,\ldots , M\}$: $I,J \neq \emptyset = I \cap J$.
$\pi (\tau)$
is a polynomial in $\tau$, of degree $2m-1$ at the $m$th critical point,
which determines the structure of the charge density in the scaling
region, given by $\frac{i}{2\pi} {\rm Disc} \!<\! \wt (\tau) \!>_{\nu=0}$
in the saddle-point approximation.
The last term in (\ref{star}) should be $\frac{1}{\sg + \tau} \int \R dt_{0}$
from the discussion following (\ref{newsm}),
which is exponentially small in $\nu$ at $t_{0} \rightarrow
-\infty$ but is $O(1)$ for $t_{0} \rightarrow \infty$. The physical
meaning of the divergence as $\tau + \sg \rightarrow 0^{+}$, is that
large loops become unsuppressed, since $\wt (\tau)$ is the
Laplace transform w.r.t. $\tau$ of the renormalised
macroscopic loop wavefunction (\ref{loo}).
Because of the ${\rm e}^{-\sg l}$ dependence of (\ref{loo}) the Laplace
transform only exists for $\tau > -\sg$.\footnote{In fact there are also
negative powers of $l$ in $w(l)$ preventing naive Laplace transformation,
but these come from low genus and can be systematically isolated.} As
indicated in the discussion following (\ref{box}), a concomitant
divergence appears in the charge density as
$\tau + \sg \rightarrow 0^{-}$.
\section{Discussion}
The previous calculations have shown that type 3 solutions yield very
simple ${\rm e}^{-\sg l}$ behaviour for macroscopic loops. This
exponential behaviour is much like the $\nu$-perturbative result. At
each order of a WKB expansion the charge density has support on the
half-line $(-\infty,-u]$ of the real $\tau$ axis, specified
in the leading approximation by the (single)
cut in $\sqrt{\pi (\tau)}$. The rest of the real $\tau$ axis describes
the loop function $\wt (\tau)$ which on inverse Laplace transformation is
seen to have ${\rm e}^{-ul}$ times power law behaviour. At genus zero
$w(l)$ contains universal terms with inverse powers of $l$, as emphasised
in refs.\cite{mss}, since $\pi (\tau)$ is polynomial:
\beq
<w(l)> = \frac{1}{\nu} \sum_{k\geq 0} k!t_{k} l^{-k-1/2} + O(l^{1/2})
\eeq
The Laplace transform does not exist because of these terms, but one
can proceed by differentiating w.r.t. $\tau$ a sufficient number of
times. It is in this sense that $w(l)$ and $\wt(\tau)$ are transforms of
one another. Alternatively one can work at fixed `area' instead of
fixed bulk cosmological constant. The offending terms are analytic in
the latter and so do not contribute to an inverse transform to fixed
`area'. They correspond to finite loops  spanned by infinitesimal
surfaces.

The $\nu$-non-perturbative exponential behaviour in $l$ of loops for
type 1 and type 2 solutions is more complicated. For type 1 the
discreteness of the spectrum of $-D^{2} + u$ implies that $<w(l)>$
behaves like an infinite sum of exponentials with different arguments
\cite{bdss}. Only in the $l \rightarrow \infty$ limit does one recover
a simple ${\rm e}^{-e_{0}l}$ behaviour, where $e_{0}$ is the lowest
eigenvalue. It is sometimes suggested that a solution satisfying a
loop equation such as (\ref{star}), derived from a path integral
representation, is `physical' and that, by implication, one that does
not is `unphysical'. There is no known path integral formula for type 1
solutions. This is not (presently) known to contradict any physical
principle however. The hermitian matrix model is the path integral
representation of type 2 solutions, and shows that the loop expectation
is always diverging as $l \rightarrow \infty$ for such solutions \cite{bmp}.
This markedly different behaviour is due to the fact that the charge
density has support on the whole spectral line, in particular it has
an exponential tail. For a tail $\rho (e) \sim {\rm exp}(-|e|^{p})$ as
$e \rightarrow -\infty$ a dimensional argument shows that $<\!w(l)\!>
\sim {\rm exp} l^{1+1/(p-1)}$ as $l \rightarrow \infty$. This means that
one cannot define the loop function $\wt (\tau)$ in this case.

To conclude one may note some possible generalisations
of the results of this letter
to other models with $c \leq 1$. For the  $(p,q)$ minimal models
described by the generalised KdV hierarchy \cite{dou}, the string equation
(scaling equation) $[\tilde{P},Q]=Q$ will provide
new solutions analoging those of type 3.
Generally one only knows how to treat macroscopic
loops embedded at a single point in the line of
$q-1$ points.
By the same argument \cite{mms}, shifting the non-derivative
part of $Q$, one can
identify a parameter coupling to a boundary operator.
Perturbatively it appears that only one parameter can be generated in this
way (e.g. for the Ising model (4,3)
it  is the boundary magnetic field),
which led the authors of ref.\cite{mms} to conclude that certain operators
could not be expressed in the KdV formalism.
However in the  case of type 3, non-isospectral evolution equations also
play a role in defining  couplings and one can imagine restricting the
$q-1$ types of charge to different half-lines i.e. the wall becomes
`time'-dependent.
The necessary argument in terms of the W-constraints to confirm or deny
the validity of this naive picture is a little involved. A detailed
account of the $[\tilde{P},Q]=Q$ version of other minimal models will be
given by the authors of ref.\cite{bill}.
At $c=1$ the picture is similar.
By placing a wall in the scaling region
given by the inverse quadratic potential \cite{new2} one has a stable
quantum mechanical system providing a non-perturbative definition of the
theory. More particularly the macroscopic loop amplitude
at time $t$ is non-perturbatively well-defined;
\bey
<\! w(l,t) \!> & =  & \int_{\sg}^{\infty}d\lambda \;
 \psi^{\dagger}(\lambda,t) {\rm e}^{-\lambda l} \psi(\lambda,t) \\
\psi (\lambda,t) & = & \int^{E_{F}} dE {\rm e}^{iEt} \Psi(E,\lambda)
\eey
where the wavefunctions $\Psi$  vanish at the wall $\lambda = \sg$
and the matrix model corresponds to the continuation to euclidean time.
It exhibits a dependence ${\rm exp} -l\sg$, at the
expense of introducing a linear term in the potential.
The extra parameter $\sg$ more generally has a continuous argument since
again it may be time-dependent.
Understanding of the possible flow structure at $c=1$ is still hazy.
Is stabilisation
by a simple wall, possibly fluctuating in position, the only quantum
mechanical problem non-perturbatively compatible with a flow structure
organising perturbation theory, akin to $c<1$?
\vspace{10mm}

{\bf Acknowledgements:} The influence of Tim Morris on the early
stages of this work cannot be overstated. I am also grateful to Clifford
Johnson for helpful information, and to Tim Morris and Andrea Pasquinnuci
for comments on the manuscript. This work was supported by an SERC
studentship and post-doctoral fellowship RFO/B/91/9033.
\vspace{10mm}

{\em Note Added}: After this letter was typed a preprint
appeared \cite{uni} where quantum mechanics on the half-line is discussed
with regard
to the requirement of unitarity of tachyon scattering at $c=1$.
\vfill
\newpage
\begin{center}
Figure Caption \\
\end{center}
\noindent Figure 1: This shows the (unique) numerical solution of type 3
for pure gravity.  The Hamiltonian $-D^{2} +u$ has continuous spectrum
down to the value of the left asymptote of the potential $u$ (zero in the
figure shown). Those eigenvalues are more-or-less directly related to the
positions of the Dyson gas charges on the spectral line, which should lie
on the positive halfline. Note that although there is a small well in $u$,
the previous identification indicates that it is too shallow to support
bound states below the continuum. A rough numerical estimate using the
proportions of the figure confirms this. As is explained in section 2,
more generally the left asymptote can be $u= \sg$ if the Dyson gas is
restricted to $[\sg,\infty)$. It is  possible that non-perturbative
subleties arise as $\sg$ turns negative e.g. through an instability in $u$
similar to that found in ref.\cite{flow}.

\vfill
\end{document}